\pdfoutput=0
\documentclass[11pt]{article}
\usepackage{axodraw}
\usepackage{epsfig}
\usepackage{amsfonts}
\usepackage{amsmath}
\usepackage{bm,bbm}
\usepackage{cite}
\allowdisplaybreaks[1]

\input pix.sty

\newcommand{\la}[1]{\label{#1}}
\newcommand{\be}{\begin{equation}}
\newcommand{\ee}{\end{equation}}
\newcommand{\ba}{\begin{eqnarray}}
\newcommand{\ea}{\end{eqnarray}}
\newcommand{\rmi}[1]{{\mbox{\scriptsize #1}}}
\newcommand{\fig}{Fig.~}
\newcommand{\figs}{Figs.~}
\newcommand{\eq}{Eq.~}
\newcommand{\eqs}{Eqs.~}
\newcommand{\se}{Sec.~}
\newcommand{\ses}{Secs.~}
\newcommand{\nr}[1]{(\ref{#1})}

\newcommand{\nn}{\nonumber \\}

\renewcommand{\vec}[1]{{\bf #1}}

\newcommand{\tfr}[2]{{\textstyle \frac{#1}{#2}\,}}

\renewcommand{\eq}{eq.~}
\renewcommand{\eqs}{eqs.~}
\renewcommand{\se}{sec.~}
\renewcommand{\ses}{secs.~}
\renewcommand{\fig}{fig.~}
\renewcommand{\figs}{figs.~}
\newcommand{\LT}{\mathbbm{L}} 
\newcommand{\PT}{\mathbbm{K}^\rmii{T}} 

\newcommand{\T}{\rmii{$T$}}

\newcommand{\mpl}{m_\rmii{Pl}}
\newcommand{\Nf}{N_{\rm f}}
\newcommand{\Nc}{N_{\rm c}}

\newcommand{\rmO}{{\mathcal{O}}}

\newcommand{\dA}{d_\rmii{A}^{ }}

\def\lsi{\raise0.3ex\hbox{$<$\kern-0.75em\raise-1.1ex\hbox{$\sim$}}}
\def\gsi{\raise0.3ex\hbox{$>$\kern-0.75em\raise-1.1ex\hbox{$\sim$}}}
\newcommand{\lsim}{\mathop{\lsi}}
\newcommand{\gsim}{\mathop{\gsi}}

\newcommand{\nB}{n_\rmii{B}}
\newcommand{\rmii}[1]{{\mbox{\tiny\rm{#1}}}}
\newcommand{\rmiii}[1]{{\mbox{\tiny{$\scriptstyle{\rm#1}$}}}}

\newcommand{\im}{\mathop{\mbox{Im}}}

\newcommand{\Tint}[1]{{\hbox{$\sum$}\!\!\!\!\!\!\!\int\,}_{\!\!\!\!\raise-0.9ex\hbox{$\scriptstyle{#1}$}}}
\newcommand{\Tinti}[1]{{{\Sigma}\!\!\!\!\raise0.3ex\hbox{$\int$}_\rmii{${#1}$}}}

\newcommand{\bi}{\begin{itemize}}
\newcommand{\ei}{\end{itemize}}
\newcommand{\hide}[1]{ }

\newcommand{\deltabar}{\raise-0.02em\hbox{$\bar{}$}\hspace*{-0.8mm}{\delta}}
\newcommand{\scat}[1]{\mbox{scat}^{ }_\rmi{$#1$}}
\newcommand{\s}[1]{s_{#1}}
\renewcommand{\P}{\mathcal{P}}
\newcommand{\K}{\mathcal{K}}
\newcommand{\cQ}{g} 
\newcommand{\cS}{\varphi} 
\newcommand{\cT}{ } 
\def\TAsc(#1,#2)(#3,#4,#5)%
{\SetWidth{2.0}\CArc(#1,#2)(#3,#4,#5)\SetWidth{1.0}}
\def\Lwidth{3}

\def\TAgl(#1,#2)(#3,#4,#5){\SetWidth{2.0}\PhotonArc(#1,#2)(#3,#4,#5){\Lwidth}%
{6.283 #3 mul 360 div #4 #5 sub #4 #5 sub mul sqrt mul Tdensity mul}%
\SetWidth{1.0}}
\def\TLgl(#1,#2)(#3,#4){\SetWidth{2.0}\Photon(#1,#2)(#3,#4){\Lwidth}
{#1 #3 sub #1 #3 sub mul #2 #4 sub #2 #4 sub mul add sqrt Tdensity mul}%
\SetWidth{1.0}}

\renewcommand{\picc}[1]{\;\parbox[c]{60pt}{\begin{picture}(60,30)(0,-10)
\SetWidth{1.0}\SetScale{1.3} #1 \end{picture}}\; }
\def\Lwidth{1.3}

\newcommand{\picu}[1]{\;\parbox[c]{70pt}{\begin{picture}(70,30)(-10,-5)
\SetWidth{1.0}\SetScale{0.65} #1 \end{picture}}\; }

\def\PhisgA{\picu{%
 \Asc(30,5)(22.3,27,72)%
 \Asc(30,5)(22.3,108,153)%
 \Asc(30,25)(22.3,207,333)%
 \COval(10,15)(2,2)(0){Black}{Black}%
 \COval(50,15)(2,2)(0){Black}{Black}%
 \Agl(30,27.3)(7,0,180)
 \Agl(30,27.3)(7,180,360)
 \Photon(-5,16.5)(8,16.5){1.5}{2}
 \Photon(-5,13.5)(8,13.5){1.5}{2}
 \Photon(52,16.5)(65,16.5){1.5}{2}
 \Photon(52,13.5)(65,13.5){1.5}{2}
 \Line(30,-2)(30,40)
}}
\def\PhigsA{\picu{%
 \Agl(30,5)(22.3,27,72)%
 \Agl(30,5)(22.3,108,153)%
 \Agl(30,25)(22.3,207,333)%
 \COval(10,15)(2,2)(0){Black}{Black}%
 \COval(50,15)(2,2)(0){Black}{Black}%
 \Agl(30,27.3)(7,0,180)%
 \Asc(30,27.3)(7,180,360)%
 \Photon(-5,16.5)(8,16.5){1.5}{2}
 \Photon(-5,13.5)(8,13.5){1.5}{2}
 \Photon(52,16.5)(65,16.5){1.5}{2}
 \Photon(52,13.5)(65,13.5){1.5}{2}
 \Line(30,-2)(30,40)
}}
\def\PhigsB{\picu{%
 \Agl(30,5)(22.3,27,90)%
 \Agl(30,5)(22.3,90,153)%
 \Agl(30,25)(22.3,207,270)%
 \Agl(30,25)(22.3,270,333)%
 \COval(10,15)(2,2)(0){Black}{Black}%
 \COval(50,15)(2,2)(0){Black}{Black}%
 \Lsc(30,2.7)(30,27.3)%
 \Photon(-5,16.5)(8,16.5){1.5}{2}
 \Photon(-5,13.5)(8,13.5){1.5}{2}
 \Photon(52,16.5)(65,16.5){1.5}{2}
 \Photon(52,13.5)(65,13.5){1.5}{2}
 \Line(18,-2)(42,32)
}}
\def\PhigsC{\picu{%
 \Agl(30,5)(22.3,27,90)%
 \Agl(30,5)(22.3,90,153)%
 \Agl(30,25)(22.3,207,270)%
 \Agl(30,25)(22.3,270,333)%
 \COval(10,15)(2,2)(0){Black}{Black}%
 \COval(50,15)(2,2)(0){Black}{Black}%
 \Lsc(30,2.7)(30,27.3)%
 \Photon(-5,16.5)(8,16.5){1.5}{2}
 \Photon(-5,13.5)(8,13.5){1.5}{2}
 \Photon(52,16.5)(65,16.5){1.5}{2}
 \Photon(52,13.5)(65,13.5){1.5}{2}
 \Line(42,-2)(18,32)
}}
\def\PhisxgA{\picu{%
 \Agl(30,5)(22.3,27,90)%
 \Asc(30,5)(22.3,90,153)%
 \Asc(30,25)(22.3,207,270)%
 \Agl(30,25)(22.3,270,333)%
 \COval(10,15)(2,2)(0){Black}{Black}%
 \COval(50,15)(2,2)(0){Black}{Black}%
 \Lgl(30,2.7)(30,27.3)%
 \Photon(-5,16.5)(8,16.5){1.5}{2}
 \Photon(-5,13.5)(8,13.5){1.5}{2}
 \Photon(52,16.5)(65,16.5){1.5}{2}
 \Photon(52,13.5)(65,13.5){1.5}{2}
 \Line(18,-2)(42,32)
}}
\def\PhisxgB{\picu{%
 \Agl(30,5)(22.3,27,90)%
 \Asc(30,5)(22.3,90,153)%
 \Asc(30,25)(22.3,207,270)%
 \Agl(30,25)(22.3,270,333)%
 \COval(10,15)(2,2)(0){Black}{Black}%
 \COval(50,15)(2,2)(0){Black}{Black}%
 \Lgl(30,2.7)(30,27.3)%
 \Photon(-5,16.5)(8,16.5){1.5}{2}
 \Photon(-5,13.5)(8,13.5){1.5}{2}
 \Photon(52,16.5)(65,16.5){1.5}{2}
 \Photon(52,13.5)(65,13.5){1.5}{2}
 \Line(42,-2)(18,32)
}}
\def\PhisxgC{\picu{%
 \Asc(30,5)(22.3,27,90)%
 \Agl(30,5)(22.3,90,153)%
 \Agl(30,25)(22.3,207,270)%
 \Asc(30,25)(22.3,270,333)%
 \COval(10,15)(2,2)(0){Black}{Black}%
 \COval(50,15)(2,2)(0){Black}{Black}%
 \Lgl(30,2.7)(30,27.3)%
 \Photon(-5,16.5)(8,16.5){1.5}{2}
 \Photon(-5,13.5)(8,13.5){1.5}{2}
 \Photon(52,16.5)(65,16.5){1.5}{2}
 \Photon(52,13.5)(65,13.5){1.5}{2}
 \Line(18,-2)(42,32)
}}
\def\PhisxgD{\picu{%
 \Asc(30,5)(22.3,27,90)%
 \Agl(30,5)(22.3,90,153)%
 \Agl(30,25)(22.3,207,270)%
 \Asc(30,25)(22.3,270,333)%
 \COval(10,15)(2,2)(0){Black}{Black}%
 \COval(50,15)(2,2)(0){Black}{Black}%
 \Lgl(30,2.7)(30,27.3)%
 \Photon(-5,16.5)(8,16.5){1.5}{2}
 \Photon(-5,13.5)(8,13.5){1.5}{2}
 \Photon(52,16.5)(65,16.5){1.5}{2}
 \Photon(52,13.5)(65,13.5){1.5}{2}
 \Line(42,-2)(18,32)
}}
\def\AmplHydro{\picu{%
 \Lsc(30,30)(50,15)%
 \Lsc(30,0)(50,15)%
 \COval(50,15)(2,2)(0){Black}{Black}%
 \Photon(52,16.5)(70,16.5){1.5}{2.5}%
 \Photon(52,13.5)(70,13.5){1.5}{2.5}
 \Line(30,-4)(30,4)
 \Line(26,0)(34,0)
 \Line(30,26)(30,34)
 \Line(26,30)(34,30)
}}
\def\AmplA{\picu{%
 \Lsc(40,22.5)(50,15)%
 \Lsc(30,15)(50,15)%
 \Lgl(30,30)(40,22.5)%
 \Agl(10,22)(30,-50,-20)%
 \Agl(10,22)(30,-10,0)%
 \COval(50,15)(2,2)(0){Black}{Black}%
 \Photon(52,16.5)(70,16.5){1.5}{2.5}%
 \Photon(52,13.5)(70,13.5){1.5}{2.5}
}}
\def\AmplB{\picu{%
 \Lgl(30,0)(50,15)%
 \Lgl(30,30)(50,15)%
 \Lsc(30,15)(40,7.5)%
 \COval(50,15)(2,2)(0){Black}{Black}%
 \Photon(52,16.5)(70,16.5){1.5}{2.5}%
 \Photon(52,13.5)(70,13.5){1.5}{2.5}
}}
\def\AmplC{\picu{%
 \Lgl(30,0)(50,15)%
 \Lgl(30,30)(50,15)%
 \Lsc(30,15)(40,22.5)%
 \COval(50,15)(2,2)(0){Black}{Black}%
 \Photon(52,16.5)(70,16.5){1.5}{2.5}%
 \Photon(52,13.5)(70,13.5){1.5}{2.5}
}}
\def\AmplD{\picu{%
 \Lsc(15,0)(30,15)%
 \Lgl(15,30)(30,15)%
 \Lgl(30,15)(50,15)%
 \Lgl(50,15)(65,0)%
 \COval(50,15)(2,2)(0){Black}{Black}%
 \Photon(51,17.5)(64,32.5){1.5}{3}%
 \Photon(53,15.5)(66,30.5){1.5}{3}
}}
\def\AmplE{\picu{%
 \Lsc(10,5)(30,5)%
 \Lgl(10,25)(30,25)%
 \Lgl(30,5)(30,25)%
 \Lgl(30,5)(50,5)%
 \COval(30,25)(2,2)(0){Black}{Black}%
 \Photon(32,26.5)(50,26.5){1.5}{3}%
 \Photon(32,23.5)(50,23.5){1.5}{3}
}}
\def\AmplF{\picu{%
 \Lsc(10,5)(30,25)%
 \Lgl(10,25)(18,17)%
 \Lgl(22,13)(30,5)%
 \Lsc(30,5)(30,25)%
 \Lgl(30,5)(50,5)%
 \COval(30,25)(2,2)(0){Black}{Black}%
 \Photon(32,26.5)(50,26.5){1.5}{3}%
 \Photon(32,23.5)(50,23.5){1.5}{3}
}}
\def\AmplG{\picu{%
 \Lgl(15,0)(30,15)%
 \Lgl(15,30)(30,15)%
 \Lsc(30,15)(50,15)%
 \Lsc(50,15)(65,0)%
 \COval(50,15)(2,2)(0){Black}{Black}%
 \Photon(51,17.5)(64,32.5){1.5}{3}%
 \Photon(53,15.5)(66,30.5){1.5}{3}
}}
\def\AmplH{\picu{%
 \Lgl(10,5)(30,5)%
 \Lgl(10,25)(30,25)%
 \Lgl(30,5)(30,25)%
 \Lsc(30,5)(50,5)%
 \COval(30,25)(2,2)(0){Black}{Black}%
 \Photon(32,26.5)(50,26.5){1.5}{3}%
 \Photon(32,23.5)(50,23.5){1.5}{3}
}}
\def\AmplI{\picu{%
 \Lgl(10,5)(30,25)%
 \Lgl(10,25)(18,17)%
 \Lgl(22,13)(30,5)%
 \Lgl(30,5)(30,25)%
 \Lsc(30,5)(50,5)%
 \COval(30,25)(2,2)(0){Black}{Black}%
 \Photon(32,26.5)(50,26.5){1.5}{3}%
 \Photon(32,23.5)(50,23.5){1.5}{3}
}}
\def\AmplJ{\picu{%
 \Lgl(35,15)(55,0)%
 \Lgl(35,15)(55,30)%
 \Lsc(15,15)(35,15)%
 \COval(45,7.5)(2,2)(0){Black}{Black}%
 \Photon(44,8.5)(55,17){1.5}{2}%
 \Photon(46,6.5)(57,15){1.5}{2}
}}
\def\AmplK{\picu{%
 \Lgl(35,15)(55,0)%
 \Lgl(35,15)(55,30)%
 \Lsc(15,15)(35,15)%
 \COval(45,22.5)(2,2)(0){Black}{Black}%
 \Photon(44,21.5)(55,13){1.5}{2}%
 \Photon(46,23.5)(57,15){1.5}{2}
}}
\def\AmplL{\picu{%
 \Lsc(35,15)(45,22.5)%
 \Lgl(45,22.5)(55,30)%
 \Lsc(15,15)(35,15)%
 \Agl(75,22)(30,200,230)%
 \Agl(75,22)(30,180,190)%
 \COval(35,15)(2,2)(0){Black}{Black}%
 \Photon(36,16.5)(55,16.5){1.5}{3}%
 \Photon(36,13.5)(55,13.5){1.5}{3}
}}

\makeatletter \@addtoreset{equation}{section} \makeatother
\renewcommand{\theequation}{\arabic{section}.\arabic{equation}}
\makeatletter
\renewcommand\section{\@startsection {section}{1}{\z@}%
                                   {-5.5ex \@plus -1ex \@minus -.2ex}
                                   {2.3ex \@plus.2ex}%
                                   {\normalfont\large\bfseries}}
\renewcommand\subsection{\@startsection{subsection}{2}{\z@}%
                                     {-3.25ex\@plus -1ex \@minus -.2ex}%
                                     {1.5ex \@plus .2ex}%
                                     {\normalfont\normalsize\bfseries}}
\renewcommand\thesection {\@arabic\c@section}
\renewcommand\thesubsection   {\thesection.\@arabic\c@subsection}
\renewcommand{\@seccntformat}[1]{%
\csname the#1\endcsname.\hspace{1.0em}}
\makeatother

\begin{document}

\flushbottom

\begin{titlepage}

\begin{flushright}
May 2022
\end{flushright}
\begin{centering}
\vfill

{\Large{\bf
 Gravitational wave background from non-Abelian reheating
  \\[3mm]
 after axion-like inflation
}} 

\vspace{0.8cm}

P.~Klose, 
M.~Laine, 
S.~Procacci

\vspace{0.8cm}

{\em
AEC, 
Institute for Theoretical Physics, 
University of Bern, \\ 
Sidlerstrasse 5, CH-3012 Bern, Switzerland \\}

\vspace*{0.8cm}

\mbox{\bf Abstract}
 
\end{centering}

\vspace*{0.3cm}
 
\noindent
A pseudoscalar inflaton $\varphi$, coupled to the topological charge 
density $F\tilde{F}$ of a non-Abelian sector, can decay to gauge 
bosons ($\varphi\to g g $), which may thermalize rapidly.
The friction felt by~$\varphi$ is then increased by  
non-Abelian ``strong sphalerons'', leading to a self-amplifying  
process that can efficiently heat up the medium. We determine a lower 
bound for the gravitational wave production rate from such a process, 
originating via hydrodynamic fluctuations and particle collisions, 
in terms of a minimal number of parameters. Only a moderate fraction 
of energy density is converted to gravitational waves, suggesting  
that non-Abelian models may avoid the overproduction observed in 
some Abelian cases.

\vfill


\end{titlepage}

\tableofcontents

%
\section{Introduction}
\la{se:intro}

The dynamics of the early universe is dictated, on one hand, 
by gravitational physics, accounting 
for a period of inflationary expansion that produced the seeds 
for structure formation. On the other hand, 
the microscopic properties of matter are governed by  
particle physics. The two descriptions 
connect to each other during reheating, in which 
the vacuum energy density, 
sustaining inflation, 
is converted to thermal radiation, 
carried by particles.  
The process establishes the highest temperature, $T^{ }_\rmi{max}$,
that can meaningfully be talked about. 

It would be interesting to understand theoretically 
the dynamics of reheating, 
and to develop empirical tests for it.  
As we are studying early moments, penetrating 
probes need to be considered, with gravitational 
waves as an obvious candidate~\cite{kt,preheat2,preheat3,preheat4}.
As tensor perturbations, 
gravitational waves are 
also produced by inflation itself~\cite{tensor}. However, 
the spectra that originate from inflation and reheating 
have different shapes, 
with the latter peaking at higher frequencies, 
all the way to the microwave range MHz...GHz. 
This could be within the reach of future observation, 
even if novel avenues need to be explored
(cf.\ ref.~\cite{uhf} for a review).  

The purpose of this paper is to study reheating within a framework
that is simple enough to be 
tractable without heavy numerics, yet rich enough to offer
for variants with different observational consequences. 

Specifically, 
we consider a pseudoscalar inflaton field $\varphi$~\cite{ai,ai_em,ai_rev}, 
whose interactions are governed by the Lagrangian 
\be
 \mathcal{L} 
 \; \supset \;
 \frac{1}{2}\, \partial^\mu\varphi\, \partial_\mu\varphi
 - V(\varphi)  
 - \frac{\varphi \,  \chi}{f^{ }_a}
 \;, 
 \quad
 \chi \;\equiv\;
 c^{ }_\chi \, 
 \epsilon^{\mu\nu\rho\sigma}_{ }
  g^2 F^{c}_{\mu\nu} F^{c}_{\rho\sigma}
 \;, 
 \quad
 c^{ }_\chi \; \equiv \; 
 \frac{1}{64\pi^2}
 \;, \la{L}
\ee
where 
$
 F^c_{\mu\nu} \equiv \partial^{ }_\mu A^c_\nu - \partial^{ }_\nu A^c_\mu
 + g f^{cde}_{ }A^d_\mu A^e_\nu
$ 
is the Yang-Mills field strength, 
$\Nc^{ }$ is the number of colours, 
$ c \in \{ 1,..., \Nc^2 - 1  \}$,
$g^2 \equiv 4\pi\alpha$ is the gauge coupling,   
and $f^{ }_a$ is the axion decay constant. 
In the context of inflation, 
the allure of this model stems from its incorporation of 
interactions with Standard Model-like gauge fields, 
but in the special way 
that they do not spoil the desired flatness 
of the potential. That said, 
\eq\nr{L} does 
involve a non-renormalizable operator, so to keep
the description self-consistent, it can only be applied
for low energy and momentum scales of $\varphi$, i.e.\ 
$\epsilon^{ }_\varphi \ll \max( 4\pi f^{ }_a, f^{ }_a / \alpha )$.

The inflationary predictions of \eq\nr{L} depend 
on the shape of the potential $V(\varphi)$.
A much-used example mimics an instanton-induced 
periodic structure, 
\be
 V(\varphi) \simeq
 m^2 f_b^2\, \Bigl[ 1 - \cos\Bigl( \frac{\varphi}{f^{ }_b} \Bigr) \Bigr]
 \;, \la{V_ai}
\ee
where on the semiclassical level
$f^{ }_b = f^{ }_a$ (this may change after renormalization).\footnote{%
 If a confinement scale $\Lambda$ related to instantons 
 is known, the parameters are 
 constrained by $m f_b^{ } \sim \Lambda^2$.
 } 
However, there are other possibilities, for example as
given in \eqs(4)--(8) of ref.~\cite{potentials}.
As we are concerned with a late stage, 
we may expand the potential around 
its global minimum ($\varphi^{ }_\rmi{min}=0$). 
This then leads to a universal shape that is also 
familiar from chaotic inflation~\cite{chaotic}, 
\be
 V(\varphi) \simeq 
 \frac{ m^2 \varphi^2 }{2}
 \;. \la{V_chaotic}
\ee

According to \eqs\nr{L} and \nr{V_chaotic}, our framework 
depends on the parameters $f^{ }_a$, $m$, $\alpha$. 
It would be easy 
to enrich the setup by introducing more parameters, 
for instance via the shape of $V(\varphi)$. 
Furthermore, if the plasma contains fermions
($\psi^{ }_k$, $k = 1,...,\Nf^{ }$), then it is natural 
to couple $\varphi$ also to the corresponding pseudoscalar operators, 
$\partial^{\mu}_{ }[\bar{\psi}^{ }_k \gamma^{ }_\mu \gamma^{ }_5 \psi^{ }_k]$
and 
$ i m^{ }_k \bar{\psi}^{ }_k \gamma^{ }_5 \psi^{ }_k$, 
where $m^{ }_k$ is a fermion mass.
The coefficient of each operator is, {\it a priori}, independent. 
If the fermions couple to the gauge fields $A^a_\mu$, 
then there is a certain redundancy in the couplings, 
as dictated by the axial anomaly equation.
At the same time, fermions affect the dynamics of the gauge sector, 
with anomalous processes generating an 
effective chemical potential for left and right-chiral modes, 
which influences the friction 
felt by $\varphi$~\cite{fermions}. 
However, in the following,  
we restrict ourselves to the minimal setup
of \eqs\nr{L} and \nr{V_chaotic}. 

Our presentation is organized as follows. 
We start with a brief review of approaches to 
reheating in \se\ref{se:mechanism}, motivating the idea that 
in the non-Abelian case the density matrix of the system can be 
parametrized with a rapidly increasing temperature-like variable. 
The main part is \se\ref{se:gravity}: after reviewing 
how gravitational waves are produced from such a system
(\se\ref{ss:general}), we compute the contributions at 
long (\se\ref{ss:hydro}) and short wavelengths (\se\ref{ss:boltzmann}), 
and illustrate the results numerically (\se\ref{ss:numerics}). 
The findings are summarized in \se\ref{se:concl}.

%
\section{Mechanism of sphaleron-induced reheating}
\la{se:mechanism}

Recently, 
a large body of literature has appeared
in which the gauge 
fields in the operator $\chi$ of \eq\nr{L} 
have Abelian nature, and contribute 
then to the gravitational wave background  
(an incomplete list can be found in 
refs.~\cite{ab01,ab1,ab11,ab111,ab12,ab121,ab1210,ab1211,ab122,ab1221,%
ab13,ab14,ab15,ab16,ab2,ab21,ab3,ab4,ab5,ab51,ab6,ab60,ab61,ab62,ab7}). 
Physically, Abelian fields could represent either a ``dark photon'', 
or the Standard Model hypercharge gauge field. 
In the following, we restrict ourselves instead to
non-Abelian gauge fields, which is arguably
more natural if one thinks of embedding the Lagrangian in some 
Grand Unified framework.\footnote{%
 The assumption of a non-zero gauge field background,
 resulting in a partial breaking 
 of a non-Abelian gauge symmetry, has led
 to a much-studied scenario (cf.\ ref.~\cite{su2orig}, and 
 refs.~\cite{su2a,su2b,su2c,su2d,su2e,su2f}
 for recent work and references). 
 Here we consider the case without any background, 
 as it breaks discrete symmetries.}

The methods employed in the above studies  
follow, roughly speaking,
two lines. One relies on the 
solution of mode equations (cf.,\ e.g.,\ ref.~\cite{ai_em}), 
the other 
on classical field theory simulations 
(cf.,\ e.g.,\ refs.~\cite{su2num1,u1num,su2num2}). 
While simulations should account for the full non-perturbative
dynamics of momentum modes with large occupation numbers, 
they are not sensitive to phenomena where
the occupation number is of order unity.
In particular,   
the quantum mechanical vacuum decays that dominate 
at early stages, or the thermalization 
towards the Bose distribution that takes place at the end, 
are not captured. 

The framework that we adopt is a different one, modelling the 
complicated dynamics of the non-Abelian sector through the assumption
that the system effectively thermalizes. 
In practice, this is closely related to the paradigm
of warm inflation~\cite{warm_old}, 
well established in the axion inflation 
context~\cite{axT00,axT0,axT1,axT2,axT25,axT3}. 
Our new implementation guarantees that the initial vacuum-like
decays are correctly incorporated as well~\cite{warm5}. 

The physical thinking behind this framework is that 
non-Abelian gauge fields tend to 
thermalize more rapidly than 
other types of interactions, as both particle number and momenta
are changed at each cubic vertex. In fact, the heat bath has 
been argued to represent an attractor solution 
even during the inflationary stage~\cite{warm6,warm7}. 
General aspects of non-Abelian thermalization have
been reviewed in ref.~\cite{therm}.

For the case of \eq\nr{L}, the friction
felt by $\varphi$ takes a special form. 
The reason is that, in the limit of low frequencies, 
the real-time topological susceptibility is rendered non-zero  
through the non-perturbative dynamics 
mediated by ``strong sphalerons''~\cite{mms}. 
Incorporating this physics, an interesting warm inflation 
scenario has been found~\cite{warm1,warm3,warm4}.  

In concrete terms, adopting the signature ($+$$-$$-$$-$), and splitting 
overall energy-momentum conservation into field and radiation 
parts~\cite{hydro1}, the equations of motion take the form 
\ba
 {\varphi^{;\mu}_{ }}^{ }_{;\mu} 
 + \Upsilon u^\mu_{ }\varphi^{ }_{;\mu} + V'(\varphi)
 & \simeq & 
 0 
 \;,  \la{eom_varphi} \\
 \bigl[(e^{ }_r + V - T\partial^{ }_\T V) u^\mu_{ } \bigr]^{ }_{;\mu}
 + (p^{ }_r - V) {u^{\mu}_{ }}^{ }_{;\mu}
 - V'(\varphi) u^\mu_{ }\varphi^{ }_{;\mu} 
 & \simeq & 
 \Upsilon
 \bigl( u^\mu_{ }\varphi^{ }_{;\mu} \bigr)^2_{ }
 \;, \la{eom_plasma} 
\ea
where 
$T$ is a local temperature; 
$u^{\mu}_{ }$ is a local plasma four-velocity; 
$\Upsilon$ is a friction coefficient; 
and $e^{ }_r$ and $p^{ }_r$
are respectively the energy density and pressure of radiation. 

The key issue is to fix $\Upsilon$, as it originates from 
the operator in \eq\nr{L}. This problem was addressed in 
ref.~\cite{warm5}, where it was shown that the value is 
determined by the response function of the medium, 
evaluated at an appropriate frequency scale, $\omega$.
Thereby both vacuum physics (for $\omega \gg \pi T$) and
sphaleron physics (for $\omega \lsim \alpha^2 T$)
are incorporated. 

For the present investigation, we can for the most part 
treat $\Upsilon$ as an independent parameter. 
Its dynamical form in terms of $f^{ }_a$, $m$, $\alpha$ and $T$ can be 
inserted at the end (cf.\ \se\ref{ss:numerics}). We also
keep in mind that, 
as demonstrated in ref.~\cite{warm5} in the context of 
the ``weak regime'' of warm inflation, 
the system can reheat up to temperatures  
$T^{ }_\rmi{max} \sim f^{ }_a / \alpha$ after inflation, 
where the influence of the operator $\chi$ in \eq\nr{L}
becomes of order unity. 

%
\section{Gravitational wave production from reheating}
\la{se:gravity}

As already alluded to, 
an important physics phenomenon 
associated with the reheating
process is the production of 
a gravitational wave background~\cite{kt,preheat2,preheat3,preheat4}. 
In the Abelian case, this can be used for 
constraining the model parameters 
(cf., e.g., ref.~\cite{potentials} and references therein).
In the current section, we study gravitational 
wave production within the framework of \se\ref{se:mechanism}.

The empirical significance of a gravitational wave background
depends on its wavelength. 
The physics of inflation corresponds to 
large wavelengths, or low frequencies, $\ll$ Hz today.
In contrast, gravitational waves produced at reheating peak
at high frequencies, which are in fact similar to those originating from 
a thermal plasma~\cite{gravity_qualitative,gravity_lo,ringwald}.

To carry out the computation, we organize it in an ``adiabatic''
approximation, assuming that plasma reactions are fast compared with
the Hubble rate ($\alpha^2 T > H$) and considering physical momenta
well within the horizon ($k \gg H$).
The first part is certainly
satisfied during and after reheating~\cite{warm5}. 
Under these conditions we can operate
in a local Minkowskian frame.\footnote{%
  Recently, the regime of typical inflationary
  momenta exiting the horizon ($k < H$) has been considered~\cite{tt}.
  The analysis of hydrodynamic fluctuations bears
  a conceptual similarity to \se\ref{ss:hydro}, however the shear
  viscosity was taken over from self-interacting
  scalar field theory~\cite{jeon}
  rather than from interactions with a heat bath.
  In principle our formalism could be generalized to apply to
  such a situation and then be used for addressing the full range
  of momenta relevant for warm inflation
  (cf.,\ e.g.,\ ref.~\cite{warm_x1} for a concrete realization),
  however the approximation of a local Minkowskian frame needs to
  be abandoned for the momentum modes
  that exit the horizon,
  even if the basic physical processes remain the same. 
}

It is generally believed 
that non-equilibrium processes 
produce more gravitational waves than 
equilibrium ones, as the latter reflect the maximal entropy
of the underlying system, so that
we should not be able to discern many features. 
Therefore thermal production as considered here is likely to 
set a {\em lower bound} for the full rate. 

%
\subsection{General features}
\la{ss:general}

In a local Minkowskian frame, 
the production
rate of the energy density carried by gravitational radiation can be 
expressed as~\cite{gravity_qualitative,gravity_lo} 
\be
 \frac{{\rm d}e^{ }_\rmiii{GW}}{{\rm d}t \, {\rm d}\ln k}
 \; = \; 
 \frac{k^4 \dot{f}^{ }_\rmiii{GW}}{\pi^2}
 \;, \la{degw}
\ee
where $f^{ }_\rmiii{GW}$ is the polarization-average phase-space
density of gravitons, and the dot stands for a time
derivative. On general grounds~\cite{db}, 
the evolution equation for $f^{ }_\rmiii{GW}$ 
takes the form
\be
 \dot{f}^{ }_\rmii{GW}
 \; = \; 
 \Gamma(k) \, 
 \bigl[ \nB^{ }(k) - f^{ }_\rmii{GW} \bigr]
 + 
 \rmO\biggl( \frac{1}{\mpl^4} \biggr)
 \;, \la{dfgw}
\ee
where $\nB^{ }(k) \equiv 1/(e^{k/T} - 1)$ is the Bose distribution, 
and $\mpl^{ } \approx 1.221 \times 10^{19}$~GeV is the Planck mass. 
In practice, 
$
 f^{ }_\rmii{GW} \ll \nB^{ }(k)
$, 
so the right-hand side can be approximated as 
$
 \Gamma(k)\, \nB^{ }(k)
$.

The dynamical information about the processes taking place is 
encoded in the interaction rate $\Gamma(k)$, which in turn can 
be expressed as~\cite{gravity_lo} 
\be
 \Gamma(k) = 
 \frac{4 \pi
 \LT_{ }^{\alpha\beta;\mu\nu} 
 \im  
 G^\rmiii{R}_{\alpha\beta;\mu\nu} (k,k)}{ k\, \mpl^2}
 \;, \la{Gamma}
\ee
where $G^\rmiii{R}_{\alpha\beta;\mu\nu}(\omega,k)$ is the retarded
correlator related to the energy-momentum tensor $T^{ }_{\mu\nu}$,
\be
 G^\rmiii{R}_{\alpha\beta;\mu\nu}(\omega,k) 
 \; \equiv \; 
 i \int_\mathcal{X} e^{i\mathcal{K}\cdot\mathcal{X}}_{ }
 \theta(t) 
 \bigl\langle\, \bigl[ 
     T^{ }_{\alpha\beta}(\mathcal{X}) \, , \, 
     T^{ }_{\mu\nu}(0) 
 \bigr] \, \bigr\rangle^{ }_\T  
 \;, \quad
 \mathcal{K}\cdot\mathcal{X} \equiv \omega t - \vec{k}\cdot\vec{x}
 \;, \la{GR}
\ee
$\langle...\rangle^{ }_\T$ denotes a thermal average, 
and $\LT^{ }$ is 
a transverse and traceless projector, 
\be
 \LT^{ }_{\alpha\beta;\mu\nu}
 \; \equiv \; 
 \frac{ 
      \PT_{\alpha\mu}\PT_{\beta\nu} 
   +  \PT_{\alpha\nu}\PT_{\beta\mu}
      }{2} 
 - 
 \frac{
      \PT_{\alpha\beta}
      \PT_{\mu\nu}
      }{D-2}
 \;, \quad
 \PT_{\alpha\beta}
 \; \equiv \; 
 \eta^{ }_{\alpha i}\eta^{ }_{\beta j}
 \biggl(
   \delta^{ }_{ij} - \frac{k^{ }_i k^{ }_j}{k^2} 
 \biggr)
 \;, \la{def_L}
\ee
with $D = 4$ denoting the spacetime dimension.
Important properties of $\PT_{ }$ are
\be
 \PT_{\alpha\mu}
 \,
 {\PT_{\beta}}_{ }^{\mu}
 = 
 - \PT_{\alpha\beta}
 \;, \quad
  {\PT_{\mu}}_{ }^{\mu}
 = 2 - D 
 \;. \la{contractions}
\ee
In practice, it is sometimes convenient to
choose the special frame in which $ \vec{k} = k\, \vec{e}^{ }_z $, 
whereby rotational symmetry implies that\footnote{%
  An algebraic way to verify the prefactor is given in footnote~2 of 
  ref.~\cite{gravity_lo}. Physically, 
  there are two transverse-traceless polarizations,
  and if we choose the one in non-diagonal components as a representative, 
  {\it viz.} $(T^{ }_{xy}+T^{ }_{yx})/\sqrt{2}$, then the equality   
  $T^{ }_{xy} = T^{ }_{yx}$ leads to the additional factor
  $(\sqrt{2})^2 = 2$. 
 }
\be
 \LT_{ }^{\alpha\beta;\mu\nu} G^\rmiii{R}_{\alpha\beta;\mu\nu}
 \; \stackrel{\rmii{$D=4$}}{=} \;
 4 G^\rmiii{R}_{xy;xy} \bigr|^{ }_{\vec{k} = k\, \vec{e}^{ }_z}
 \;. \la{projection}
\ee
 
The method to compute $\im G^\rmiii{R}_{xy;xy}$ 
depends on the momentum range considered. 
For very small momenta, 
we find ourselves in the so-called hydrodynamic domain. 
Then
$\im G^\rmiii{R}_{xy;xy}$ evaluates to $\eta \omega$, where 
$\eta$ is the shear viscosity~\cite{gravity_qualitative}. The task
therefore is to determine the contribution of the inflaton
field $\varphi$ to the shear
viscosity, a topic that we address in \se\ref{ss:hydro}. In contrast, 
for typical thermal momenta, parametrically $k\sim \pi T$, elementary
gauge bosons and axions 
can be resolved as quasi-particles. Then we are faced with
a Boltzmann type of a computation, 
which is presented in \se\ref{ss:boltzmann}.

%
\subsection{Hydrodynamic domain}
\la{ss:hydro}

We start by computing the correlator in \eq\nr{projection} 
in the hydrodynamic domain,
i.e.\  at very small frequencies and momenta.\footnote{%
  The scales at which hydrodynamics applies can be 
  estimated as follows. Consider a sound wave, carrying the energy 
  $\omega^{ } = c^{ }_s k$, where the speed of sound is 
  $c^{ }_s \simeq 1/\sqrt{3}$. It is damped by viscous effects, 
  with the rate $\Gamma^{ }_s \sim \eta k^2 / (e + p)$, 
  where $e + p = T s$ is the enthalpy and $s$ the 
  entropy density. Shear viscosity is parametrically of order 
  $\eta \sim T^3/\alpha^2$. Hydrodynamics is applicable
  for $\Gamma^{ }_s \ll \omega^{ }$, which converts
  to $ \omega, k \ll \alpha^2 T $.
 }  
In this regime, 
elementary particles cannot be resolved, 
and the degrees of freedom relevant for describing 
the gauge plasma are hydrodynamic fluctuations.
The contribution of a weakly coupled scalar field to 
\eq\nr{projection} has been
determined in ref.~\cite{fluct} in this domain, 
and here we improve upon that computation, 
by ameliorating its ultraviolet (UV) sensitivity.

In the hydrodynamic domain, the traceless part of the 
energy-momentum tensor reads
\be
 T^{ }_{\mu\nu} 
 \,\supset\,  
 \partial^{ }_\mu \varphi\, \partial^{ }_\nu \varphi
 + T^{\,r}_{\mu\nu}
 \;, \la{Tmunu_hydro}
\ee
where $T^{\,r}_{\mu\nu}$ is the contribution of radiation. 
Total energy-momentum is conserved, with the coefficient $\Upsilon$
extracting energy from $\varphi$, according to \eq\nr{eom_varphi}, 
and transmitting it to the plasma, according to \eq\nr{eom_plasma}. 
For us the information needed about this dynamics is the retarded
propagator of $\varphi$, determined in ref.~\cite{fluct}. In the
local rest frame, it takes the form 
\be
 \Pi^\rmii{R}_{ }(\omega,\vec{p}) 
 = 
 \frac{1}{-\omega^2 + \epsilon_p^2 - i \omega \Upsilon}
 \;, \quad
 \epsilon_p^2 \; \equiv \; p^2 + m_{\cT}^2 
 \;. \la{PiR}
\ee

In order to now compute \eq\nr{projection}, 
we make use of the real-time formalism of thermal field theory, 
in the so-called Keldysh ($r/a$) basis (cf.,\ e.g.,\ ref.~\cite{review}). 
Then the propagator becomes a matrix, 
\be
 \left( 
 \begin{array}{cc}
   G^{ }_{rr} & G^{ }_{ra} \\ 
   G^{ }_{ar} & G^{ }_{aa} 
 \end{array}
 \right) 
 \; = \;
 \left( 
 \begin{array}{cc}
   \Delta & -i \Pi^\rmiii{R}_{ } \\ 
   -i \Pi^\rmiii{A}_{ } & 0 
 \end{array}
 \right) 
 \;, \la{prop}
\ee
where all components are determined by \eq\nr{PiR}, 
\be
 \Delta = [1 + 2\nB^{ }(\omega)] \rho
 \;, \quad
 \rho = \im \Pi^\rmii{R}_{ } = 
 \frac{ \omega \Upsilon }
      {(\omega^2_{ } - \epsilon_p^2)^2 + \omega^2\Upsilon^2}
 \;, \quad
 \Pi^\rmii{A}_{ } = (\Pi^\rmii{R}_{ })^*_{ }
 \;. \la{relations}
\ee

As for the vertices, they are obtained by substituting 
$\varphi^{ }_1 = \varphi^{ }_r + \varphi^{ }_a/2$ and
$\varphi^{ }_2 = \varphi^{ }_r - \varphi^{ }_a/2$
in the Lagrangian 
$
 \mathcal{L}(\varphi^{ }_1) - \mathcal{L}(\varphi^{ }_2)
$.
Applying this first to how a metric perturbation $h$ couples
to the energy-momentum tensor $T$, we obtain
\be
 h^{ }_1 T^{ }_1 - h^{ }_2 T^{ }_2 = h^{ }_a T^{ }_r + h^{ }_r T^{ }_a
 \;. \la{haTr_1}
\ee
Repeating the same with the specific structure from \eq\nr{Tmunu_hydro} yields
\be
 h^{ }_1\, \varphi^{ }_{1,x} \varphi^{ }_{1,y} -
 h^{ }_2\, \varphi^{ }_{2,x} \varphi^{ }_{2,y} 
 = 
 h^{ }_a 
 \, \Bigl[ 
 \underbrace{ 
             \varphi^{ }_{r,x}\varphi^{ }_{r,y} 
           + \frac{\varphi^{ }_{a,x}\varphi^{ }_{a,y}}{4}
            }_{ T^{ }_r }
 \Bigr]
 + 
 h^{ }_r 
 \, \Bigl[ 
 \underbrace{
              \varphi^{ }_{a,x}\varphi^{ }_{r,y} + 
              \varphi^{ }_{r,x}\varphi^{ }_{a,y}
            }_{ T^{ }_a }
 \Bigr]
 \;. \la{haTr_2}
\ee

We can now consider the retarded correlator of the energy-momentum tensor.
Given that the propagator $G^{ }_{aa}$ vanishes, the contribution
of $\varphi$ originates from 
\be
 -i G^\rmii{R}_{xy;xy}  \bigr|^{ }_{\vec{k} = k\, \vec{e}^{ }_z}
   = \langle T^{ }_r T^{ }_a \rangle
   = \bigl\langle 
      (  \varphi^{ }_{r,x}\varphi^{ }_{r,y}  )
      (  \varphi^{ }_{a,x}\varphi^{ }_{r,y} + 
         \varphi^{ }_{r,x}\varphi^{ }_{a,y}  )
     \bigr\rangle
 \;. 
\ee
Carrying out the Wick contractions, going over to momentum space, 
inserting \eqs\nr{prop} and \nr{relations}, taking the imaginary part, 
and symmetrizing
in $\mathcal{P}^{ }_1\leftrightarrow \mathcal{P}^{ }_2$, yields
\ba
 && \hspace*{-2.5cm} 
 \im G^\rmii{R}_{xy;xy} (\omega,k)
 \bigr|^{ }_{\vec{k} = k\, \vec{e}^{ }_z}
  \; = \;
 \int^{ }_{\mathcal{P}^{ }_1,\mathcal{P}^{ }_2}
 \deltabar(\mathcal{K} - \mathcal{P}^{ }_1 - \mathcal{P}^{ }_2)
 \bigl[ 1 + \nB^{ }(\omega^{ }_1) + \nB^{ }(\omega^{ }_2) \bigr]
 \nn 
 & \times & 
 \bigl\{ 
   \rho^{ }_{,x,x}(\mathcal{P}^{ }_1)
   \rho^{ }_{,y,y}(\mathcal{P}^{ }_2)
  + 
   \rho^{ }_{,y,y}(\mathcal{P}^{ }_1)
   \rho^{ }_{,x,x}(\mathcal{P}^{ }_2)
  + 
      2 
      \rho^{ }_{,x,y}(\mathcal{P}^{ }_1)
      \rho^{ }_{,y,x}(\mathcal{P}^{ }_2)
 \bigr\} 
 \;, \la{step1}
\ea
where
$
 \int^{ }_{\mathcal{P}}\,\deltabar(\mathcal{P}) \equiv 1
$.

%
\begin{figure}[t]
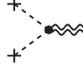


\begin{eqnarray*}
&& \hspace*{-1.3cm}
 \AmplHydro 
\end{eqnarray*}

\vspace*{-3mm}

\caption[a]{\small 
 The physical process responsible for \eq\nr{step2}, 
 leading to the gravitational wave production rate shown 
 in \eq\nr{step5}. 
 Dashed lines denote the inflaton $\varphi$; 
 a doubled line a graviton; 
 a blob the operator $T^{ }_{\mu\nu}$; 
 and crosses thermal fluctuations,
 which are transmitted to $\varphi$
 via the coefficient $\Upsilon$.  
} 
\la{fig:hydro}
\end{figure}
%

In order to evaluate \eq\nr{step1}, 
it is helpful to factorize the Bose distributions as 
$
 1 + \nB^{ }(\omega^{ }_1) + \nB^{ }(\omega^{ }_2) 
 = \nB^{-1}(\omega) 
   \nB^{ }(\omega^{ }_1) 
   \nB^{ }(\omega^{ }_2)
$. 
Integrating over $\mathcal{P}^{ }_2$,  
denoting $\vec{p} \equiv \vec{p}^{ }_1$, 
and partial fractioning the $\omega^{ }_1$-dependence 
of the spectral functions, we then obtain
\ba
 && \hspace*{-2.5cm} 
 \im G^\rmii{R}_{xy;xy} (\omega,k)
 \bigr|^{ }_{\vec{k} = k\, \vec{e}^{ }_z}
  \; = \;
  \nB^{-1}(\omega) \int_\vec{p} 
  \frac{ p_x^2 p_y^2 }
  {4\tilde{\epsilon}_p^{ }\tilde{\epsilon}_{pk}^{ }}
 \int_{-\infty}^{\infty} 
 \! \frac{{\rm d}\omega^{ }_1}{2\pi} 
 \nn
 & \times & 
 \nB^{ }(\omega^{ }_1)
 \biggl[
   \frac{\Upsilon}
        {(\omega^{ }_1 - \tilde{\epsilon}_p^{ })^2 + \frac{\Upsilon^2}{4} } 
   - 
   \frac{\Upsilon}
        {(\omega^{ }_1 + \tilde{\epsilon}_p^{ })^2 + \frac{\Upsilon^2}{4} } 
 \biggr]
 \nn 
 & \times & 
 \nB^{ }(\omega - \omega^{ }_1)
 \biggl[
   \frac{\Upsilon}
        {(\omega - \omega^{ }_1 - \tilde{\epsilon}_{pk}^{ })^2
        + \frac{\Upsilon^2}{4} } 
   - 
   \frac{\Upsilon}
        {(\omega - \omega^{ }_1 + \tilde{\epsilon}_{pk}^{ })^2
     + \frac{\Upsilon^2}{4} } 
 \biggr]
 \;, \la{step2}
\ea
where 
$
 \tilde{\epsilon}_p^{\,2} \equiv \epsilon_p^2 - \Upsilon^2/4
$, 
$
 \tilde{\epsilon}_{pk}^{\,2} \equiv \epsilon_{pk}^2 - \Upsilon^2/4
$,
and 
$
 \epsilon_{pk}^2 \equiv (\vec{p-k})^2 + m^2
$. 
The corresponding physical process is depicted 
in \fig\ref{fig:hydro}.

As a next step, we may integrate over $\omega^{ }_1$. 
This can be done with the residue theorem. We first note that 
the Bose distribution $\nB^{ }(\omega^{ }_1)$ has poles at 
$\omega^{ }_1 = i \omega^{ }_n \equiv  i 2\pi n T$, 
$n\in \mathbbm{Z}$.  
The pole at $\omega^{ }_1 = 0$ is lifted by the expression in 
the square brackets, and the same is true for the pole 
at $\omega - \omega^{ }_1 = 0$, 
from $\nB^{ }(\omega - \omega^{ }_1)$. 
If we close the contour in the upper half-plane, 
the remaining contributions are from  
$\omega^{ }_1 \in 
\{ i \omega^{ }_n ,   
\omega + i \omega^{ }_n, 
\pm \tilde{\epsilon}^{ }_p + i \Upsilon/2 , 
\pm \tilde{\epsilon}^{ }_{pk} + \omega + i \Upsilon/2
\} 
$, with $n\ge 1$.

Even if the residues are readily determined and the corresponding 
expression could be integrated numerically, it is helpful to 
put it in a more transparent form, by considering 
\be 
 \omega,k,\Upsilon \ll {\epsilon}^{ }_p \sim \pi T
  \;. \la{ir_domain}
\ee
The magnitude of $\epsilon^{ }_p$ originates from looking at the domain
where the $\vec{p}$-integrand becomes suppressed, 
because of factors
$\sim 1/(\epsilon^{ }_p \pm i \omega^{ }_n)^m$ or 
$\sim \nB^{ }(\epsilon^{ }_p)$, with $m \in \mathbbm{N}^+_{ }$.
The inequality part of \eq\nr{ir_domain} 
is certainly true, given that in the hydrodynamic 
domain, $\omega,k \lsim \alpha^2 T$,
and that for the maximal
temperature, {\it viz.}\ $T^{ }_\rmi{max} \sim f^{ }_a / \alpha$, 
$\Upsilon^{ }_\rmi{max} \sim \alpha^5 T^3_\rmi{max} / f_a^2 
                        \sim \alpha^3 T^{ }_\rmi{max}$.

Now, the contributions originating from 
$\omega_1^{ } \in\{ i\omega^{ }_n,\omega + i \omega^{ }_n \}$
are strongly suppressed in the limit of \eq\nr{ir_domain}, 
being parametrically $\sim \omega \Upsilon^2 T$. The other 
residues yield much larger contributions, 
parametrically $\sim \omega T^4 / \Upsilon$.
In the domain of \eq\nr{ir_domain} we can furthermore approximate
\be
 \tilde{\epsilon}^{ }_p \approx \epsilon^{ }_p 
 \;, \quad
 \tilde{\epsilon}^{ }_{pk} \approx \epsilon^{ }_p - v^{ }_z k
 \;, \quad
 v^{ }_z \; \equiv \; \frac{p^{ }_z}{\epsilon^{ }_p}
 \;. 
\ee
Then the leading contributions combine into 
\be
 \im G^\rmii{R}_{xy;xy} (\omega,k)
 \; \stackrel{\omega,\,k,\Upsilon \ll \pi T}{\approx} \; 
 \nB^{-1}(\omega) \Upsilon  
 \int_\vec{p} 
 \frac{ p_x^2 p_y^2 
 \,\nB^{ }(\epsilon^{ }_p)
 \bigl[
  1 +  \nB^{ }(\epsilon^{ }_p) 
 \bigr]
      }{{\epsilon}_p^2\, 
        [(v^{ }_z k + \omega)^2 + \Upsilon^2]} 
 \;. \la{step4}
\ee 
This illustrates a crossover from the regime $\omega,k \gg \Upsilon$,
where the result is suppressed by $\Upsilon$, to that at 
$\omega,k \ll \Upsilon$, where the result is enhanced by $1/\Upsilon$.
The physical reason for why 
the gravitational wave production rate
at very low frequencies (or the shear viscosity), 
is inversely proportional to the coupling~\cite{jeon},
is that the most weakly interacting particle species 
display the strongest hydrodynamic fluctuations.

The angular integral in \eq\nr{step4} can be carried out, by going over
to spherical coordinates. Inserting 
$
 \int_0^{2\pi} \! {\rm d}\phi \, 
 \cos^2\!\phi\,\sin^2\!\phi = \pi / 4 
$
and denoting 
\be
 \mathcal{F}(\omega,vk,\Upsilon)
 \; \equiv \; 
 \int_{-1}^{+1} \! {\rm d}z \, 
 \frac{(1-z^2)^2}{(v k z + \omega)^2+\Upsilon^2}
 \;, \quad v \;\equiv\; \frac{p}{\epsilon^{ }_p}
 \;, 
\ee
which at light-cone $\omega = k$ has the limiting values\footnote{%
 The full expression reads
 $
 \mathcal{F}(\omega,vk,\Upsilon)
  =  \frac{2(9 \omega^2 - 5 v^2k^2 - 3 \Upsilon^2)}{3 v^4 k^4}
  - 
 \frac{2\omega(\omega^2 - v^2k^2 - \Upsilon^2)}{v^5k^5}
 \ln\bigl[ 
 \frac{(\omega + vk)^2 + \Upsilon^2}
      {(\omega - vk)^2 + \Upsilon^2}
 \bigr] 
 + 
 \frac{(\omega^2 - v^2 k^2)^2 - 
 2 (3 \omega^2 - v^2 k^2)\Upsilon^2 + \Upsilon^4}{v^5k^5\Upsilon} 
 \bigl[
  \arctan\bigl( \frac{\omega + v k}{\Upsilon} \bigr) 
 - 
  \arctan\bigl( \frac{\omega - v k}{\Upsilon} \bigr) 
 \bigr]
 $.
} 
\be
 \mathcal{F}(k,vk,\Upsilon)
 \;\approx\; 
 \left\{ 
 \begin{array}{ll}
  \displaystyle 
  \frac{16}{15\Upsilon^2}\;, & 
  k \ll \Upsilon \\[2mm] 
  \displaystyle 
  \frac{4}{3 v^5 k^2}
  \biggl[
   2 v(3 - 2 v^2) - 3(1-v^2) 
  \ln\frac{1+v}{1-v}   
  \biggr]
  \; \stackrel{v\approx 1}{\approx} \;
  \frac{8}{3k^2}
  \;, & 
   k \gg \Upsilon
 \end{array}
 \right. 
 \;, \la{calF_appro}  
\ee
the result from 
\eqs\nr{degw}--\nr{Gamma} and \nr{projection}
can be expressed as 
\be
 \frac{{\rm d}e^{ }_\rmiii{GW}}{{\rm d}t \, {\rm d}\ln k}
 \;\stackrel{k,\Upsilon \ll \pi T}{\approx}\; 
 \frac{k^3 \Upsilon }{2\pi^3\mpl^2}
 \int_0^{\infty} \!  
 \frac{ {\rm d}p \,p^6_{ } 
 \,\nB^{ }(\epsilon^{ }_p)
 \bigl[
  1 +  \nB^{ }(\epsilon^{ }_p) 
 \bigr]
      }{{\epsilon}_p^2} 
 \, 
 \mathcal{F}  
 \biggl( k, \frac{p k}{\epsilon^{ }_p},\Upsilon \biggr)
 \;. \la{step5}
\ee
For the largest wavelengths, where $\mathcal{F}$ can be 
approximated by the first line of \eq\nr{calF_appro}, 
this is illustrated numerically in \fig\ref{fig:spectra}(middle).

When we go from $k \ll \Upsilon$ to $k \gg \Upsilon$, 
\eqs\nr{calF_appro} and \nr{step5} indicate that the growth of 
$
 {{\rm d}e^{ }_\rmiii{GW}}/({{\rm d}t \, {\rm d}\ln k})
$ 
moderates from $\sim k^3$ into $\sim k$. But the function is still
growing, and in fact most of the 
energy density carried by gravitational waves lies 
at larger momenta, $k \sim \pi T$. 
We now turn to how this dominant contribution can be determined. 

%
\subsection{Boltzmann domain}
\la{ss:boltzmann} 

%
\begin{figure}[t]
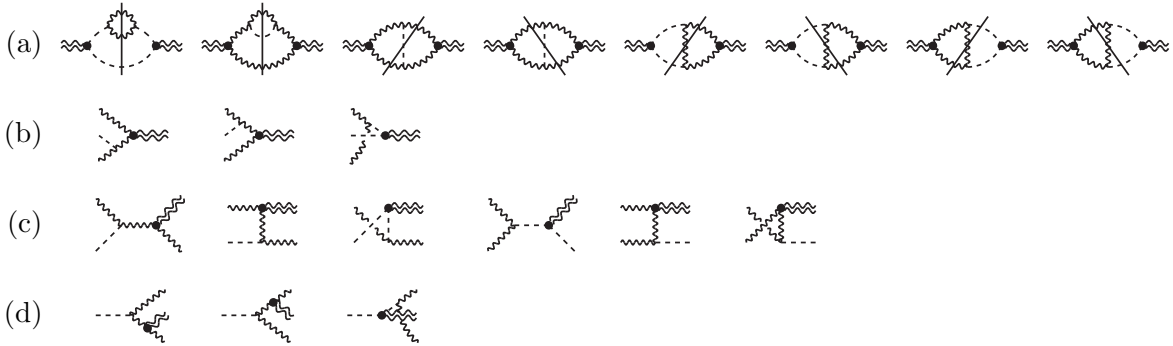


\begin{eqnarray*}
\mbox{(a)} && \hspace*{-0.8cm}
 \PhisgA  \hspace*{-0.8cm}
 \PhigsA  \hspace*{-0.8cm} 
 \PhigsB  \hspace*{-0.8cm} 
 \PhigsC  \hspace*{-0.8cm} 
 \PhisxgA \hspace*{-0.8cm} 
 \PhisxgB \hspace*{-0.8cm} 
 \PhisxgC \hspace*{-0.8cm} 
 \PhisxgD \\ 
\mbox{(b)} && \hspace*{-1.1cm}
 \AmplB  \hspace*{-1.0cm}
 \AmplC  \hspace*{-1.0cm} 
 \AmplA   \\
\mbox{(c)} && \hspace*{-0.8cm}
 \AmplD  \hspace*{-0.8cm}
 \AmplE  \hspace*{-1.0cm} 
 \AmplF  \hspace*{-1.0cm}
 \AmplG  \hspace*{-0.8cm}
 \AmplH  \hspace*{-1.0cm} 
 \AmplI   \\ 
\mbox{(d)} && \hspace*{-0.8cm}
 \AmplJ  \hspace*{-1.0cm}
 \AmplK  \hspace*{-1.0cm} 
 \AmplL   
\end{eqnarray*}

\vspace*{-3mm}

\caption[a]{\small 
(a)~matrix elements squared contributing to gravitational wave production
in the Boltzmann domain, 
represented as ``cuts'' of a 2-point correlator of the energy-momentum
tensor. 
Dashed lines denote the inflaton $\varphi$; 
wiggly lines gauge fields; 
doubled lines gravitons; 
blobs the operator $T^{ }_{\mu\nu}$; 
(b)~the corresponding $3\to 1$ amplitudes, 
which are not kinematically allowed, 
but can be used for deriving \eq\nr{Theta};  
(c)~$2\leftrightarrow 2$ processes, obtained by crossing symmetries
from the set~(b);  
(d)~likewise, for kinematically permitted $1\to 3$ decays.
} 
\la{fig:boltzmann}
\end{figure}
%

At larger momenta, elementary particle excitations can be resolved, 
and we need to consider the microscopic form of the energy-momentum tensor. 
Omitting trace parts, which drop out when projected
with \eq\nr{def_L}, 
the fields appearing in \eq\nr{L} give the contribution 
\be 
 T^{ }_{\mu\nu}
 \,\supset\,
 \partial^{ }_\mu \varphi\, \partial^{ }_\nu \varphi
  - F^c_{\mu\alpha} {F^c_{\nu}}_{ }^{\alpha}
\ee
to the traceless part.
These components couple to the propagating part of
the graviton field ($h$).
We are interested in the contribution to graviton production
that involves one appearance of the vertex in \eq\nr{L}, 
as the processes without this vertex were already 
considered in ref.~\cite{gravity_lo}.  
Various processes are depicted in \fig\ref{fig:boltzmann}.
(The $2\to 1$ channel $\cS\cS\to h$ 
is not kinematically allowed with on-shell particle states.) 

A way to represent and evaluate the rates of the reactions
in \fig\ref{fig:boltzmann} has been presented in 
ref.~\cite{phasespace}.
In the following, we adopt its methods and notation. 
The procedure starts by considering 
the processes in \fig\ref{fig:boltzmann}(b), 
which are not kinematically allowed, but have a simple would-be 
algebraic structure, as the non-equilibrium particle and 
the plasma particles are on different sides of the reaction. 
This contribution is represented as 
\be
 \LT_{ }^{\alpha\beta;\mu\nu} 
 \im  
 G^\rmiii{R}_{\alpha\beta;\mu\nu} (k,k)
 \; \supset \;
 \scat{1 \to 3}(\cQ^{ }_1,\cS,\cQ^{ }_3)
 \,\Theta(\P^{ }_{\cQ_1},\P^{ }_\cS,\P^{ }_{\cQ_3})
 \;, \la{scat1to3} 
\ee
where $\scat{1 \to 3}$ 
is a phase-space average,\footnote{%
 In ref.~\cite{phasespace},
 the non-equilibrium particle
 was defined to be the initial state, 
 i.e.\ time was running in the opposite direction, 
 which explains 
 the reference to a $1\to 3$ process. In the particle
 production language of \fig\ref{fig:boltzmann}(b), 
 it is more intuitive to depict the non-equilibrium particle 
 as a final state, yielding a $3\to 1$ reaction. 
}
and $\cQ^{ }_1$, $\cQ^{ }_3$ label 
two (identical) gauge bosons. 

The dynamical information concerning the production process
enters through the function~$\Theta$ in \eq\nr{scat1to3}, 
which may be referred to as ``matrix element squared''. 
More precisely, if we couple $T^{ }_{\mu\nu}$ 
to polarization vectors $ \hat{h}_{ }^{\mu\nu} $ of a would-be source
($\mathcal{L}\supset \hat{h}_{ }^{\mu\nu} T^{ }_{\mu\nu}$), 
and replace the sum over the polarizations through 
\eq\nr{def_L}, {\it viz.} 
\be
 \sum_\lambda \hat{h}^{\alpha\beta}_{(\lambda)} 
              \hat{h}^{\mu\nu *}_{(\lambda)}
 \;\equiv\; \LT^{\alpha\beta;\mu\nu}_{ }
 \;, 
\ee
then 
\be
 \Theta(\P^{ }_{\cQ_1},\P^{ }_\cS,\P^{ }_{\cQ_3})
 \; = \; \frac{1}{2} \sum_{\lambda, s_1,s_3} 
 \bigl|\, 
   \mathcal{M}^{ }_{\hat{h}_{(\lambda)}\to g_{(s_1)}\,\varphi\, g_{(s_3)}}
 \, \bigr|^2_{ }
 \;, \la{Theta}
\ee
where $s^{ }_{1},s^{ }_{3}$ label the helicities  
of the final-state gauge bosons, and $\tfr12$ accounts 
for the gauge bosons being identical (with this factor we can integrate
over the full phase space without the danger 
of overcounting).\footnote{%
 Alternatively, $\Theta$ can be extracted from 
 the definition in \eq\nr{scat1to3}, 
 i.e.\ by computing the retarded 
 2-point correlator of the energy-momentum tensor and taking its 
 imaginary part, as illustrated in \fig\ref{fig:boltzmann}(a).
 Either way, it is important to crosscheck 
 the gauge independence of the result. 
 } 

The tensor $\LT^{\alpha\beta;\mu\nu}_{ }$, defined in \eq\nr{def_L}, 
contains a quadratic appearance of the projector~$\PT_{ }$. 
After contractions with the metric tensor, 
linear and zeroth order terms in $\PT_{ }$ 
can appear as well (cf.\ \eq\nr{contractions}). 
Redundancies can be eliminated by making use of the fact
that $\PT_{ }$ is orthogonal to $\K$, implying
\be 
 \PT_{\alpha\beta} \P_{1}^\alpha \P_{2}^\beta 
 = 
%
 \PT_{\alpha\beta} \, 
 \frac{
    \P_{3}^\alpha \P_{3}^\beta  
  - \P_{1}^\alpha \P_{1}^\beta  
  - \P_{2}^\alpha \P_{2}^\beta  
  }{2} 
 \;, \quad \mbox{for} \quad
 \K = \P^{ }_1 + \P^{ }_2 + \P^{ }_3
 \;, 
\ee
with analogous relations obtained through the 
relabellings $1\leftrightarrow 3$ and $2\leftrightarrow 3$. 

\newcommand{\gamm}{\mu} 
\newcommand{\lamb}{\nu} 
\newcommand{\kapp}{\rho} 
\newcommand{\delt}{\sigma} 
\newcommand{\tgamm}{\bar\mu} 
\newcommand{\tlamb}{\bar\nu} 
\newcommand{\tkapp}{\bar\rho} 
\newcommand{\tdelt}{\bar\sigma} 

A further ingredient of the computation is the Levi-Civita tensor
$\epsilon^{\mu\nu\hspace*{-0.3mm}\rho\sigma}_{ }$ from \eq\nr{L}. 
In the matrix elements
squared, it appears in the structure
\ba
 && \hspace*{-1.0cm} 
 \epsilon^{\gamm\lamb\hspace*{-0.3mm}\kapp\delt}_{ }
 \epsilon^{\tgamm\tdelt\hspace*{-0.3mm}\tkapp\tlamb}_{ }
 A^{ }_{\gamm}
 B^{ }_{\kapp}
 C^{ }_{\tgamm}
 D^{ }_{\tkapp}
 \\
 & = & 
 \bigl(A\cdot C\, B\cdot D - A\cdot D\, B\cdot C \bigr)
 \bigl( \eta^{\lamb\tlamb}_{ } \eta^{\delt\tdelt}_{ }
       -\eta^{\lamb\tdelt}_{ } \eta^{\delt\tlamb}_{ }\bigr)
   + 
   \bigl( A^{\tlamb}_{ } B^{\tdelt}_{ } - A^{\tdelt}_{ } B^{\tlamb}_{ } \bigr)
   \bigl( C^{\lamb}_{ } D^{\delt}_{ } - C^{\delt}_{ }D^{\lamb}_{ } \bigr)
 \nn 
 & + &
   \bigl( A\cdot D\, B^{\tdelt}_{ } - B\cdot D\, A^{\tdelt}_{ } \bigr) 
   \bigl(\eta^{\lamb\tlamb}_{ } C^{\delt}_{ } - 
         \eta^{\delt\tlamb}_{ } C^{\lamb}_{ } \bigr)
   +
   \bigl( A\cdot D\, B^{\tlamb}_{ } - B\cdot D\, A^{\tlamb}_{ } \bigr) 
   \bigl(\eta^{\delt\tdelt}_{ } C^{\lamb}_{ } - 
         \eta^{\lamb\tdelt}_{ } C^{\delt}_{ } \bigr)
 \nn 
 & + &
   \bigl( B\cdot C\, A^{\tdelt}_{ } - A\cdot C\, B^{\tdelt}_{ } \bigr) 
   \bigl(\eta^{\lamb\tlamb}_{ } D^{\delt}_{ } - 
         \eta^{\delt\tlamb}_{ } D^{\lamb}_{ } \bigr)
   +
   \bigl( B\cdot C\, A^{\tlamb}_{ } - A\cdot C\, B^{\tlamb}_{ } \bigr) 
   \bigl(\eta^{\delt\tdelt}_{ } D^{\lamb}_{ } - 
         \eta^{\lamb\tdelt}_{ } D^{\delt}_{ } \bigr)
 \;. 
 \nonumber 
\ea
The remaining contractions 
are a bit lengthy but conveniently handled, e.g.,\ 
with FORM~\cite{form}.

After these steps, and denoting 
$\s{12}^{ }\equiv (\P^{ }_{\cQ^{ }_1} + \P^{ }_\cS)^2$, 
$\s{13}^{ }\equiv (\P^{ }_{\cQ^{ }_1} + \P^{ }_{\cQ^{ }_3})^2$, 
$\s{23}^{ }\equiv (\P^{ }_\cS + \P^{ }_{\cQ^{ }_3})^2$, 
the result can be expressed as 
\ba
 \Theta(\P^{ }_{\cQ^{ }_1},\P^{ }_\cS,\P^{ }_{\cQ^{ }_3}) 
 & = & 
 \frac{16 g^4 \dA c_\chi^2}{f_a^2}
 \biggl\{\,
 (\s{13}^{ } - m_{\cT}^2)^2 
 \nn
 & - & 
 4 
 m_{\cT}^2 (\s{13}^{ } - m_{\cT}^2) \,
 \biggl(
   \frac{p_{1\perp}^2}{\s{23}^{ }}
 + \frac{p_{2\perp}^2}{\s{13}^{ } - m_{\cT} ^2}
 + \frac{p_{3\perp}^2}{\s{12}^{ }} 
 \biggr)
 \nn 
 & + & 
 2
 m_{\cT}^4  \,
 \biggl(
   \frac{p_{1\perp}^2}{\s{23}^{ }}
 + \frac{p_{2\perp}^2}{\s{13}^{ } - m_{\cT} ^2}
 + \frac{p_{3\perp}^2}{\s{12}^{ }} 
 \biggr)^2_{ }
 \nn 
 & + & 
 2
 m_{\cT}^4  \,
 \biggl[
 \frac{1}{\s{12}^{ }\s{23}^{ }} 
 + 
 \frac{1}{\s{12}^{ }(\s{13}^{ } - m_{\cT}^2 )} 
 + 
 \frac{1}{\s{23}^{ }(\s{13}^{ } - m_{\cT}^2 )}   
 \biggr]
 \nn
 &  & \quad \times \,  
 \Bigl[
     p_{1\perp}^4 
  +  p_{2\perp}^4
  +  p_{3\perp}^4
  - 2 \bigl(
        p_{1\perp}^2 p_{2\perp}^2
      + p_{1\perp}^2 p_{3\perp}^2
      + p_{2\perp}^2 p_{3\perp}^2 
      \bigr) 
 \Bigr]
 \,\biggr\}
 \;, \la{Theta_long} \hspace*{6mm} 
\ea
where $\dA \equiv \Nc^2 - 1$, 
and we have denoted 
\be
 p_\perp^2  
 \; \equiv \; 
 \PT_{\alpha\beta} \P_{ }^\alpha \P_{ }^\beta 
 \;.
\ee

For massless gravitons and gauge bosons, 
the kinematic invariants are constrained by 
$
 \s{12}^{ } + \s{13}^{ } + \s{23}^{ } = m_{\cT}^2 
$, 
leading to a remarkable simplification of \eq\nr{Theta_long}.
Indeed, the last term of \eq\nr{Theta_long} is eliminated by
\be
 \frac{1}{\s{12}^{ }\s{23}^{ }} 
 + 
 \frac{1}{\s{12}^{ }(\s{13}^{ } - m_{\cT}^2 )} 
 + 
 \frac{1}{\s{23}^{ }(\s{13}^{ } - m_{\cT}^2 )}  
  = 0 
 \;, 
\ee
whereas the 2nd and 3rd terms contain the combination\footnote{%
 To verify this relation, the terms in the denominator can be written as 
 $
  \s{13}^{ } - m_{\cT}^2 = 
  (\P^{ }_1 + \P^{ }_3)^2 - m_{\cT}^2 = 
  (\K^{ } - \P^{ }_2)^2  - m_{\cT}^2 \stackrel{\omega = k}{=} 
  -2 (k \epsilon^{ }_2 - \vec{k}\cdot\vec{p}^{ }_2)
 $, 
 and similarly in the other cases. 
 In the numerator, 
 $ 
  p_{2\perp}^2 = p_2^2 - (\vec{k}\cdot\vec{p}^{ }_2)^2 / k^2 
               = - m^2 + 
                 (k \epsilon^{ }_2 - \vec{k}\cdot\vec{p}^{ }_2)
                 (k \epsilon^{ }_2 + \vec{k}\cdot\vec{p}^{ }_2) / k^2
 $. 
 The latter term partly cancels against the denominator, leaving over
 $ - \sum_i[ k \epsilon^{ }_i + \vec{k}\cdot\vec{p}^{ }_i] / (2 k^2) 
  = - 1 
 $, where we made use of energy-momentum conservation. 
 }
\be
   \frac{p_{1\perp}^2}{\s{23}^{ }}
 + \frac{p_{2\perp}^2}{\s{13}^{ } - m_{\cT}^2 }
 + \frac{p_{3\perp}^2}{\s{12}^{ }}
 = - \frac{\s{13}^{ }}{\s{13}^{ } - m_{\cT}^2 }
 \;.
\ee
Inserting these simplifications, the result becomes
\be
 \Theta(\P^{ }_{\cQ^{ }_1},\P^{ }_\cS,\P^{ }_{\cQ^{ }_3}) 
  = 
 \frac{16 g^4 \dA c_\chi^2}{f_a^2}
 \,
 \frac{\s{13}^4 + m_{\cT}^8}{ \bigl( \s{13}^{ } - m_{\cT}^2 \bigr)^2}    
 \;. \la{Theta_res}
\ee
In total, \eqs\nr{degw}--\nr{Gamma} and \nr{scat1to3} yield
\be
 \frac{{\rm d}e^{ }_\rmiii{GW}}{{\rm d}t \, {\rm d}\ln k}
 \;\stackrel{k \sim \pi T}{\approx}\; 
 \frac{4 k^3 \nB^{ }(k)}{\pi\mpl^2}
 \times  
  \scat{2\leftrightarrow 2,1\leftrightarrow 3}(\cQ^{ }_1,\cS,\cQ^{ }_3)
 \,
  \Theta(\P^{ }_{\cQ^{ }_1},\P^{ }_\cS,\P^{ }_{\cQ^{ }_3})
 \;, \la{de_res}
\ee
where 
$ 
 \scat{2\leftrightarrow 2,1\leftrightarrow 3}
$ 
contains all crossed channels, 
as given in \eq(2.12) of ref.~\cite{phasespace}. 

The numerical integration of \eq\nr{de_res} can be carried out by
modifying the algorithm provided in ref.~\cite{phasespace}. 
We note that the production
peaks at temperatures $T \gg m$, and thus appearances of $m$
can be omitted in practice. In this limit there 
are no poles in \eq\nr{Theta_res}, whereby the virtual
corrections that were discussed in ref.~\cite{phasespace} 
play no role. 

%
\subsection{Numerical estimates}
\la{ss:numerics}

The purpose of this section is to 
summarize the parametric forms of the 
results that were obtained in \ses\ref{ss:hydro} and \ref{ss:boltzmann}, 
and to illustrate the corresponding prefactors numerically.

If we set $k\sim \pi T$, 
where the production rate proportional to $k^3 \nB^{ }(k)$ peaks
(cf.\ \eq\nr{de_res}), then 
$
 \frac{{\rm d}e^{ }_\rmiii{GW}}{{\rm d}t \, {\rm d}\ln k}
 \sim 
 \alpha T^7 / \mpl^2
$
for the Standard Model contribution
to gravitational wave production~\cite{gravity_qualitative}. 
The result in \eq\nr{Theta_res} contains the prefactor $\alpha^2/f_a^2$, 
implying that axion-like inflation leads to the
additional contribution 
$
 \frac{{\rm d}e^{ }_\rmiii{GW}}{{\rm d}t \, {\rm d}\ln k}
 \sim 
 \alpha^2 T^9 / ( \mpl^2 f_a^2 )
$. 
The numerical coefficients associated with these parametric
behaviours are illustrated in \figs\ref{fig:spectra}(left) and (right),
respectively, where
the running of $\alpha$ has been taken into account, 
assuming $\Nc^{ }= 3$ and a QCD-like initial value at low energies
(numerically, $\alpha\sim 0.015$ in the temperature range shown). 
In addition we have plotted the estimate in the extreme 
hydrodynamic domain, 
from \eq\nr{step5}, 
in \fig\ref{fig:spectra}(middle), even though we do not think that
this result has significance for our main conclusions. 
The hydrodynamic prediction scales as 
$
 \frac{{\rm d}e^{ }_\rmiii{GW}}{{\rm d}t \, {\rm d}\ln k}
 \sim 
 f_a^2  k^3 T^2 / ( \alpha^5 \mpl^2 )
$, 
assuming $\Upsilon \simeq \kappa \alpha^5 T^3/f_a^2$, 
with a numerical coefficient $\kappa\simeq 100$~\cite{mt}.

\begin{figure}[t]

\hspace*{-0.1cm}
\centerline{%
 \epsfysize=5.0cm\epsfbox{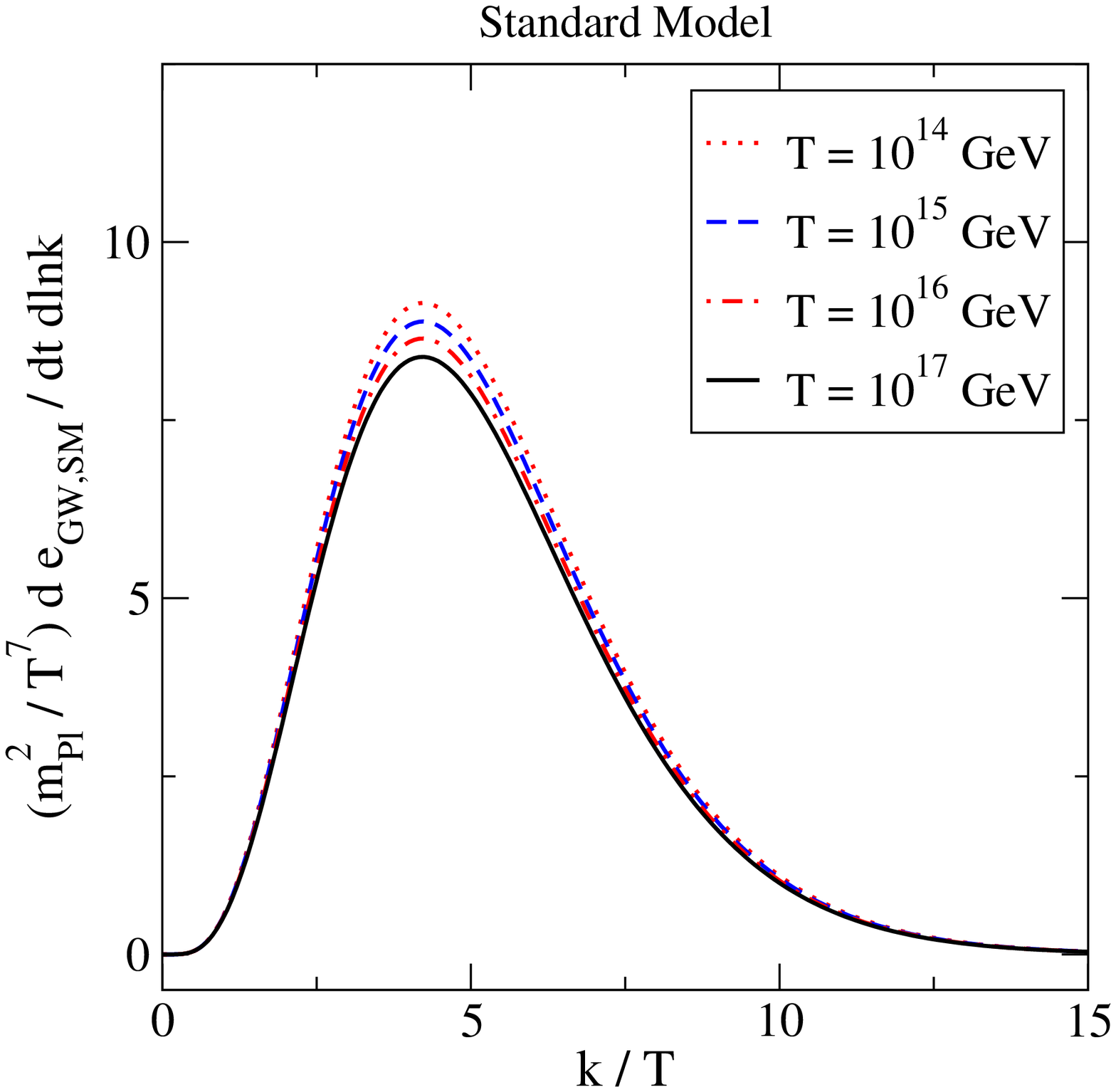}%
 \hspace{0.1cm}%
 \epsfysize=5.0cm\epsfbox{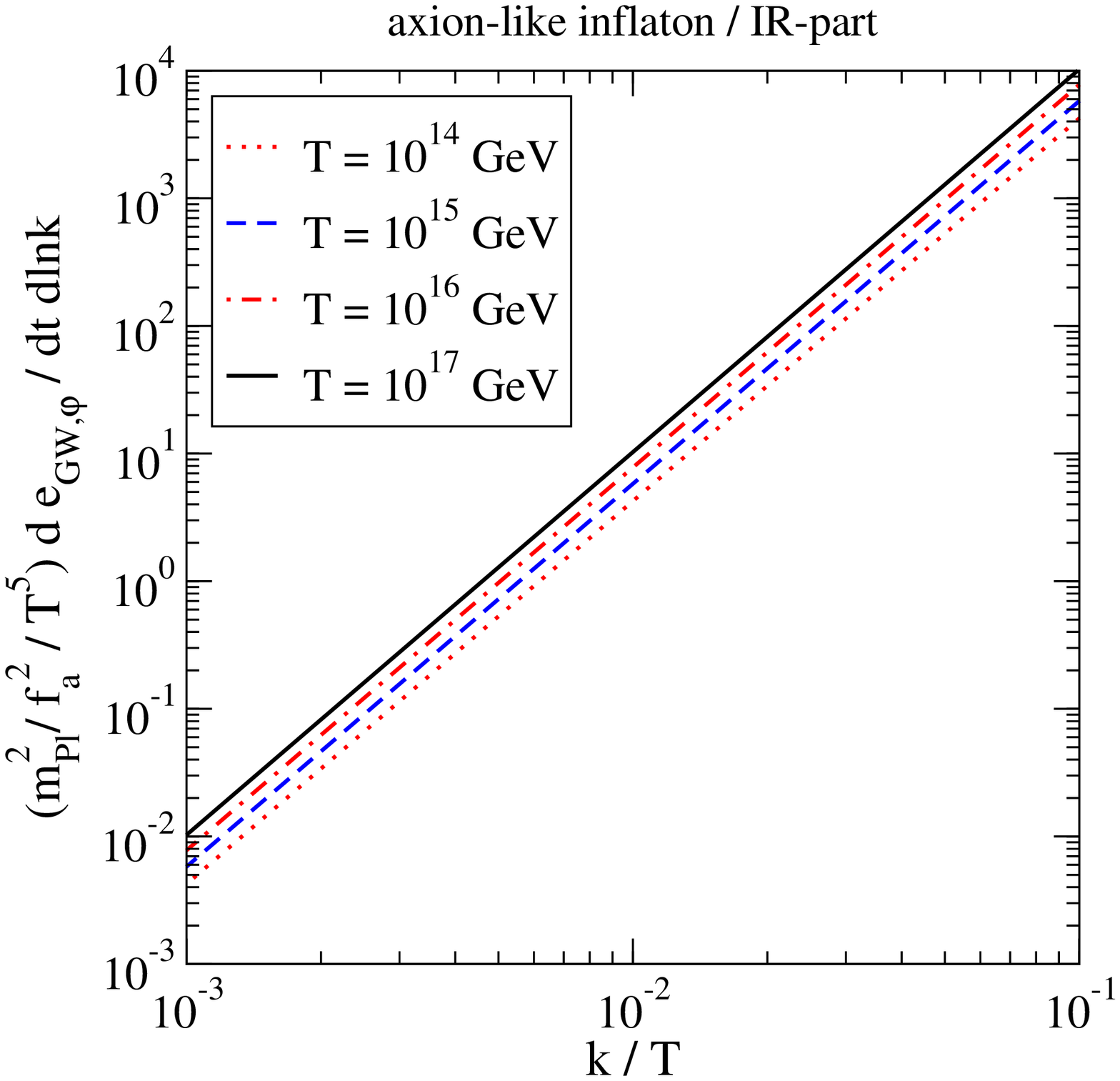}%
 \hspace{0.1cm}%
 \epsfysize=5.0cm\epsfbox{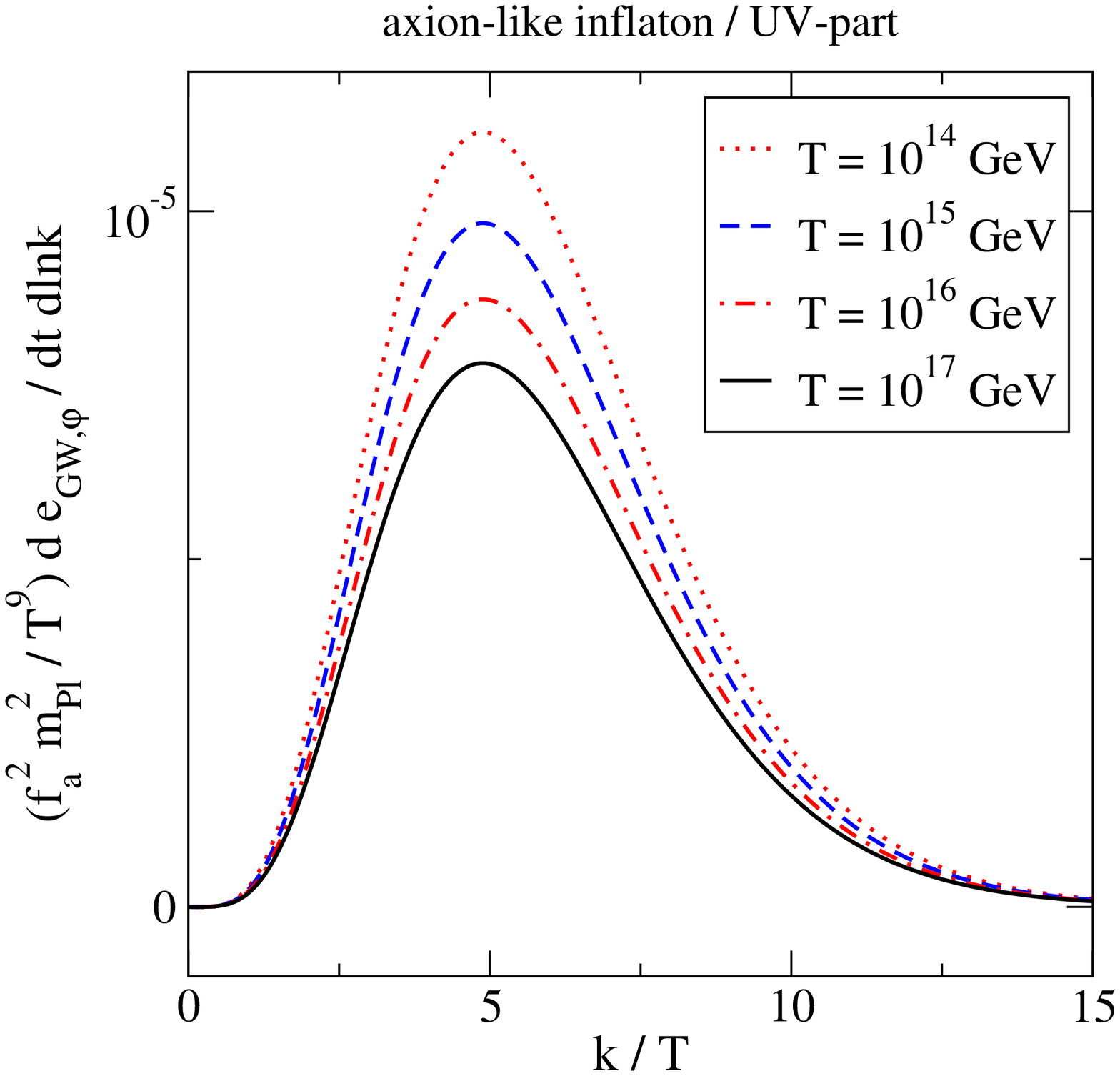}%
}

\caption[a]{\small
  Left: 
  the Standard Model contribution to the production
  rate of the energy density carried by gravitational radiation, 
  from ref.~\cite{gravity_lo},  
  normalized as 
  $
   ({m_\rmii{Pl}^2} / { T^7 })  
   \frac{{\rm d}e^{ }_\rmiii{GW}}{{\rm d}t \, {\rm d}\ln k}
  $. 
  Middle: 
  the infrared (IR) part of the axion contribution, 
  from \eq\nr{step5}, 
  normalized as 
  $
   ({m_\rmii{Pl}^2} / f_a^2  / { T^5 })  
   \frac{{\rm d}e^{ }_\rmiii{GW}}{{\rm d}t \, {\rm d}\ln k}
  $.
  The axion mass has been set to $m^{ }_{\cT} \ll T$, 
  and $\alpha$ has been set to a QCD-like value.
  Right: 
  the ultraviolet (UV) part of the axion contribution, 
  from \eqs\nr{Theta_res} and \nr{de_res}, 
  normalized as 
  $
   (f_a^2 {m_\rmii{Pl}^2} / { T^9 })  
   \frac{{\rm d}e^{ }_\rmiii{GW}}{{\rm d}t \, {\rm d}\ln k}
  $.
}

\la{fig:spectra}
\end{figure}

Given the similar shapes but different normalizations
in \figs\ref{fig:spectra}(left) and (right), 
we may expect the axion contribution to gravitational wave production
to exceed the 
Standard Model one at $T \gsim 10^3_{ } f^{ }_a $.
However, the numerical solutions 
in ref.~\cite{warm5} only reached 
$
 T^{ }_\rmi{max} \sim 200 f^{ }_a 
$. 
It thus appears that the axion contribution 
does not exceed the Standard Model one.
Furthermore, for the Standard Model contribution, 
$T^{ }_\rmi{max} = 2\times 10^{17}_{ }$~GeV  
increases the massless degrees of freedom only by
$\Delta N^{ }_\rmi{eff} \approx 10^{-3}_{ }$~\cite{gravity_lo}, 
which is very demanding to observe~\cite{ringwald}. Therefore, 
reheating through a coupling between an axion-like 
inflaton and non-Abelian gauge fields is not excluded at present,
and represents a viable scenario for the foreseeable future. 

%
\section{Summary and conclusions}
\la{se:concl}

Depending on its magnitude, 
the gravitational wave background 
produced by a reheating process~\cite{kt,preheat2,preheat3,preheat4} 
can lead to one of two possible consequences. 
If the background is substantial, 
this would be exciting as a motivation for  
possible future experiments (cf.,\ e.g., ref.~\cite{uhf}). 
If it
is moderate, we can be confident that the model
is not already excluded, as could happen in the case of  
an axion-like inflaton coupled 
to Abelian gauge fields 
(cf.,\ e.g.,\ ref.~\cite{potentials}).

For non-Abelian reheating after axion-like inflation, 
the magnitude of the gravitational wave production rate 
depends on the parameters $\alpha$, $f^{ }_a$, $m$
and $T^{ }_\rmi{max}$ (cf.\ \eq\nr{L}). 
The dependence on~$m$ is small
enough to be negligible in practice, 
provided that $m \ll \pi T^{ }_\rmi{max}$, 
as is the case towards the end of the reheating period~\cite{warm5}. 
Within the setup of \eq\nr{L}, and in the domain where most of the 
energy density lies, the dependence on $f^{ }_a$ is 
given by the power law~$1/f^{2}_a$, and
the dependence on $T^{ }_\rmi{max}$ by dimensional analysis.
Therefore the main task
has been to sort out the dependence on $\alpha$, 
and to determine the associated prefactor. 

Our numerical 
results are illustrated in \fig\ref{fig:spectra}.
Parametrically, the axion contribution exceeds 
the Standard Model for $T > f^{ }_a / \sqrt{\alpha}$.
Numerically, this has turned into $T > 10^3 f^{ }_a $, 
which is unlikely to be reached according to ref.~\cite{warm5}.
A main reason for the numerical suppression is 
the small factor~$c^{ }_\chi$ in \eq\nr{L}, which appears quadratically
in the production rate. 
If the production rate does not  
exceed the Standard Model one, it is 
not strongly constrained in the temperature range that is 
associated with normal inflationary scenarios,   
$T^{ }_\rmi{max} \ll 10^{17}$~GeV, given that in this range 
the Standard Model contribution increases the energy density
as parametrized by massless degrees of freedom only by 
$\Delta N^{ }_\rmi{eff} \ll 10^{-3}_{ }$~\cite{gravity_lo}. 

It may be wondered why the non-Abelian case differs so 
notably from the Abelian one, where an efficient tachyonic 
instability has been claimed to convert a significant fraction
of energy density to gravitational waves. The reason is that 
backreaction effects lead arguably to rapid thermalization. In a thermal
system, as we have assumed to be the case, 
tensor modes are excited only through interactions, 
whereby their production is suppressed by~$\alpha^2$.

To sharpen our conclusions, it would be nice to 
fix $T^{ }_\rmi{max}$ in terms of 
$f^{ }_a$, $m$ and $\alpha$ such that inflationary predictions are
in line with observation.
This requires going beyond the 
universal \eq\nr{V_chaotic}, 
by defining $V(\varphi)$ away from the minimum. 
Ultimately, 
it would also be great to probe non-equilibrium effects, 
and to employ a UV complete description, 
as reheating easily takes us to 
a domain where the non-renormalizable operator in 
\eq\nr{L} is having a substantial influence. 
We hope to return to some of these issues in the future.  

%
\section*{Acknowledgements}

We thank Greg Jackson for helpful discussions. 
This work was partly supported by the Swiss National Science Foundation
(SNSF) under grant 200020B-188712.

%
\appendix
\renewcommand{\thesection}{Appendix~\Alph{section}}
\renewcommand{\thesubsection}{\Alph{section}.\arabic{subsection}}
\renewcommand{\theequation}{\Alph{section}.\arabic{equation}}

\small{
%

}

\end{document}